\documentclass[journal]{IEEEtran}

\ifCLASSINFOpdf
\else
   \usepackage[dvips]{graphicx}
\fi
\usepackage{url}

\usepackage{graphicx}
\usepackage{amsmath}
\usepackage{amssymb}
\usepackage{booktabs}

\usepackage{threeparttable}
\usepackage{multirow}
\usepackage{cite}

\usepackage[pagebackref,breaklinks,colorlinks]{hyperref}

\usepackage[capitalize]{cleveref}
\crefname{section}{Sec.}{Secs.}
\Crefname{section}{Section}{Sections}
\Crefname{table}{Table}{Tables}
\crefname{table}{Tab.}{Tabs.}

\newcommand{\etal}{\textit{et al}.}

\begin{document}

\title{Audio-aware Query-enhanced Transformer \\ for Audio-Visual Segmentation}

\author{Jinxiang Liu, Chen Ju, Chaofan Ma, 
Yanfeng Wang, \IEEEmembership{Member, IEEE},\\ Yu Wang,
\IEEEmembership{Member, IEEE},  and Ya Zhang, \IEEEmembership{Member, IEEE}
\vspace{-15pt}
\thanks{The authors are with the Cooperative Medianet Innovation Center, Shanghai Jiao Tong University, Shanghai 200240, China (e-mail: \{jinxliu,  ju\_chen, chaofanma, wangyanfeng, yuwangsjtu, ya\_zhang\}\@sjtu.edu.cn).}

}

\markboth{Journal of \LaTeX\ Class Files, Vol. 14, No. 8, August 2015}
{Shell \MakeLowercase{\textit{et al.}}: Bare Demo of IEEEtran.cls for IEEE Journals}
\maketitle

\begin{abstract}
The goal of the audio-visual segmentation (AVS) task is to segment the sounding objects in the video frames using audio cues. 
However, current fusion-based methods have the performance limitations due to the small receptive field of convolution and inadequate fusion of audio-visual features. 
To overcome these issues, we propose a novel \textbf{Au}dio-aware query-enhanced \textbf{TR}ansformer (AuTR) to tackle the task. 
Unlike existing methods, our approach introduces a multimodal transformer architecture that enables deep fusion and aggregation of audio-visual features. 
Furthermore, we devise an audio-aware query-enhanced transformer decoder that explicitly helps the model focus on the segmentation of the pinpointed sounding objects based on audio signals, while disregarding silent yet salient objects.
Experimental results show that our method outperforms previous methods and demonstrates better generalization ability in multi-sound and open-set scenarios.
\end{abstract}


\IEEEpeerreviewmaketitle

\section{Introduction}\label{sec:intro}
\IEEEPARstart{A}{udio} visual segmentation (AVS) task aims to localise and segment the sounding objects in the video frames.
Previous methods~\cite{arandjelovic2018objects,afouras2020self,hu2020discriminative,pu2020active,hu2021class,hu2019deep,lin2021unsupervised,chen2021localizing,liu2022exploiting,senocak2018learning,qian2020multiple,song2022self,sun2023learning} usually tackle this problem in a self-supervised manner by using audio-visual correspondence.
However these methods are usually trained with only instance-level supervision therefore inevitably the predicted segmentations are rather coarse.
To tackle this, Zhou~\etal~\cite{zhou2022avs} present a dataset with pixel-level segmentation annotations for sounding objects in the video frames.
Benefiting from the pixel-wise supervision provided by the data, the outputs of AVS task are more precise which supports its applications in video surveillance, saliency detection and multi-modal video editing.

\begin{figure}[t]
\centering
\vspace{0.1cm}
\includegraphics[width=0.97\linewidth]{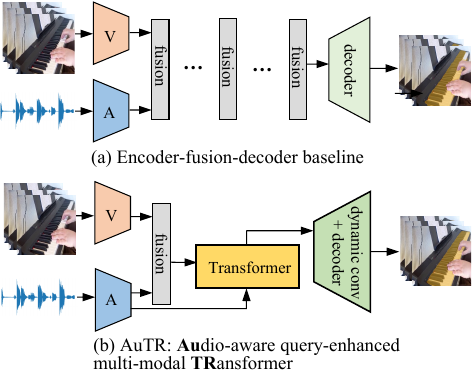}
\caption{The architecture comparison between: (a) the encoder-fusion-decoder baseline and (b) our proposed audio-aware query-enhanced transformer.}
\vspace{-0.4cm}
\label{fig:teaser1}
\end{figure}

To achieve high AVS performance, one important element is the fusion of visual and audio modalities. Following methods in other segmentation problems~\cite{ronneberger2015u,wang2022pvt,feng2021bidirectional}, Zhou \etal \cite{zhou2022avs} propose an encoder-fusion-decoder model with sophisticated fusion steps, as shown in ~\cref{fig:teaser1}(a).
The key component is the temporal pixel-wise audio-visual interaction (TPAVI) module which focuses on the fusion of the audio and visual features with convolutions.
 However, the small receptive field of convolution \cite{wang2018non,xia2022vision} makes it challenging to model the long-range dependency of visual context for the audios. Furthermore, TPAVI only performs the fusion of audio and visual features of different scales separately, without exploiting the deep guidance of audio features on the multi-modal fusion. 
 As a result, inadequate integration of audio may lead to inaccurate segmentation of target objects, even segmenting other distracting objects in some complex scenarios. 

In this letter, we propose an end-to-end Audio-aware query-enhanced TRansformer (AuTR) for the AVS task inspired by the success of the transformer architecture in multi-modal learning~\cite{xu2023multimodal}. 
AuTR is built upon a multi-modal transformer and aims to effectively fuse features inter- and intra-modalities with the powerful long-range modeling ability of attention~\cite{vaswani2017attention}. In contrast to previous methods~\cite{su2019vl, kamath2021mdetr} for fusing multi-modal inputs, we enhance audio-awareness for the transformer decoder by adopting audio-aware learnable queries. This explicitly boosts the guidance of audio information in audio-visual fusion, enabling the model to focus on segmenting objects corresponding to audio signals while ignoring other distracting objects.
Instead of directly generating target masks with the transformer decoder, we leverage dynamic convolution to predict segmentations for all queries. This allows the incorporation of more instance-specific characteristics into the model. 
Extensive experimental results demonstrate the superiority of the proposed method.

The main contributions of this letter are as follows:
\begin{itemize}
    \item We propose a novel multi-modal transformer framework to address the AVS problem, which is highly advantageous over the existing encoder-fusion-decoder scheme.
    \item To achieve accurate segmentation of target sounding objects while suppressing silent objects, the framework explicitly utilizes audio embeddings as queries for the transformer decoder to attend to sounding object-related features and guide the parameter estimation of dynamic convolution kernels for mask prediction.
   \item Experiments demonstrate the superior performance and better generalization ability of the proposed method.

\end{itemize}

\begin{figure*}[hbt]
\centering
\vspace{0.1cm}
\includegraphics[width=0.97\linewidth]{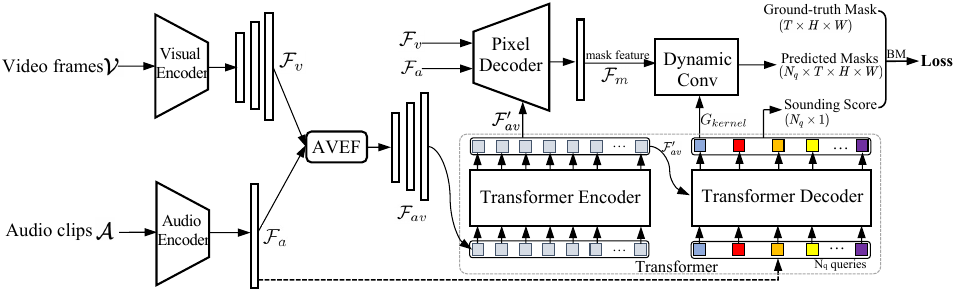}
\caption{The pipeline of the proposed audio-aware query-enhanced multi-modal transformer (AuTR). 
``BM" denotes \textit{bipartite matching} used in training stage.}
\vspace{-0.5cm}
\label{fig:framework}
\end{figure*}

\section{Method}\label{sec:method}

\subsection{Formulation \& Overview}
\noindent \textbf{Problem Definition.}
For the AVS task, 
the input data consists of a sequence of video frames, $\mathcal{V}=\left\{v_i\right\}_{i=1}^T$, where $v_{i} \in \mathbb{R}^{3 \times H_{0} \times W_{0}}$, and audios $\mathcal{A}=\left\{a_i\right\}_{i=1}^T$, 
where $a_{i} \in \mathbb{R}^{H_a\times W_a}$ denotes the audio spectrograms. 
The objective of AVS is to segment the sounding objects in each frame in $\mathcal{V}$ based on the acoustic cues. 
The target segmentation can be binary masks for each frame $\mathcal{M}=\left\{M_i\right\}_{i=1}^T$, 
where $M_{i} \in  \{0,1\}^{{H_{0} \times W_{0}}}$.

% \vspace{0.1cm}
\noindent \textbf{Framework Overview.}
As depicted in~\cref{fig:framework}, the framework contains the following main components: visual encoder to extract visual features, audio encoder to extract audio features, an audio-aware multi-modal transformer, a pixel decoder to generate mask features by aggregating the multi-modal features, and a dynamic convolution module whose parameters are generated by transformer decoder for final mask prediction.

\subsection{Feature Extraction}
First we employ pre-trained backbones to extract deep semantic features for both modalities from the input video frames and audios.

\vspace{2pt}
\noindent \textbf{Visual Encoder.} We use pretrained visual encoder to extract compact dense visual features of the video frames. 
The visual encoder can be ResNet50 \cite{he2016deep},  or transformer-based model such as Pyramid Vision Transformer (PVT) \cite{wang2022pvt}. 
To exploit features with different semantic levels, we extract three scales of visual features, denoted as $\mathcal{F}_{v}=\left\{[f^t_{v1},f^t_{v2},f^t_{v3}]\right\}_{t=1}^{T}$.

\vspace{2pt}
\noindent \textbf{Audio Encoder.} Given the audio clips corresponding to video frames $\mathcal{A}=\left\{a_i\right\}_{i=1}^T$, 
we pass them to the VGGish \cite{hershey2017cnn} model pretrained on AudioSet~\cite{gemmeke2017audio} dataset and pool the features into a vector embedding.
The obtained audio features are denoted as
$\mathcal{F}_{a}=\left\{f^t_{a}\right\}_{t=1}^{T}$, 
where $f^t_{a} \in \mathbb{R}^{C_a}$.

\subsection{Audio-Visual Encoder Fusion}
After obtaining the visual and audio features, 
we perform audio visual feature fusion that encourages the modality information interaction at the early stage.
We first adopt MLP and $1 \times 1$ conv to adjust the dimensions of audio features to $C_{av}$ and visual features $\mathcal{F}_{v}$ to the same dimension $C_{av}$ separately.
Thereafter, in AVEF module, we employ multi-head attention~\cite{vaswani2017attention} between audio and visual features at each scale, with the queries coming from visual features while the audio features serving as keys and values.
This multi-modal fusion between audio features and visual feature maps are conducted at each visual scale separately, 
ending up with multi-modal feature maps
$\mathcal{F}_{av}=\left\{[f^t_{1},f^t_{2},f^t_{3}]\right\}_{t=1}^{T}$, where $f_{\ell}^{t} \in \mathbb{R}^{C_{av} \times H_{\ell} \times W_{\ell} }$, $\ell\in \{1,2,3\}$.

\subsection{Audio-aware Multi-modal Transformer}
\noindent \textbf{Multi-modal Encoder.} The proposed audio-aware query-enhanced transformer accepts the audio-visual cross-modal feature maps $\mathcal{F}_{av}$ as inputs.
First we add fixed 2D position encoding to the feature maps. 
Then before passing the features $\mathcal{F}_{av}$ into transformer encoder, we integrate the temporal dimension to batch dimension and collapse the spatial dimensions to one dimension, to promote the transformer to process each video frame features as a sequence independently and efficiently.
Through the attention mechanism within the transformer encoder, we obtain the more global and deeply fused cross-modal features $\mathcal{F}^{\prime}_{av}$.

\noindent \textbf{Audio-aware Decoder.}
The transformer decoder aims to pinpoint the sounding object corresponding to the given audio with the cross-modal features, while ignoring the non-sounding objects and background.
To this end, we introduce a fixed number of \textit{audio-aware queries} for the decoder.
Concretely, we employ $N_q$ queries for the transformer decoder layers; these queries are initialized with the extracted audio embeddings $\mathcal{F}_a$ supplemented with learnable position embeddings.
Thus the decoder queries dependent on the audios will be guided to focus on audio-related contents rather than non-sounding objects and backgrounds.
Besides the queries, the patchfied output embeddings of the transformer encoder $\mathcal{F}^{\prime}_{av}$ are separately projected into \textit{key} and \textit{value} for the transformer decoder layers.
In this process, the transformer decoder transforms the $N$ audio-aware learnable queries along with the multi-modal features to output embeddings for all query instances.

\subsection{Mask Prediction}
 To obtain the target mask predictions, we leverage the dynamic convolution~\cite{tian2020conditional,wang2020solov2} that encodes more instance-specific characteristics to perform segmentation instead of directly predicting the masks.
Specifically, we first deploy a pixel decoder to aggregate the audio-visual multi-modal features into mask features, then use the dynamic convolutional block to predict the segmentation masks based on the mask features, detailed as follows.

\vspace{3pt}
\noindent \textbf{Pixel Decoder.} 
We first construct a light-weight pixel decoder to obtain the mask features based on the FPN~\cite{lin2017feature} architecture.
The pixel decoder accepts the multi-scale visual features $\mathcal{F}_{v}$ from visual backbone, audio features $\mathcal{F}_{a}$ from audio backbone, and the multi-modal audio-visual embeddings $\mathcal{F}^{\prime}_{av}$ from the transformer encoder;
then the decoder with cross-attention and pooling operations aggregates these features as the final enriched multi-modal mask features $\mathcal{F}_m$ for segmentation, where $\mathcal{F}_\text{m} = \{f_m^t\}_{t=1}^T, f_m^t \in \mathbb{R}^{C_m\times H_m \times W_m}$.

\vspace{3pt}
\noindent \textbf{Dynamic Conv.} The dynamic convolutional block depends on the query outputs to generate the convolution kernel weights for decoding the mask features into target masks.
We denote the output sequence for all query instances from the transformer decoder as $\mathcal{F}_Q = \{f_q^i\}_{i=1}^{N_q}$, where $f_q^i\in \mathbb{R}^D$.
We employ two-layer MLPs to predict a series of weight parameters of convolutional kernels ${G_{kernel}}$ controlled by $\mathcal{F}_Q$.
The generated kernels are:
\begin{equation}
    G_ {\text{kernel}}(\mathcal{F}_Q) = \{k_i\}_{i=1}^{N_q}, k_t \in \mathbb{R}^{D_s}.
\end{equation}

Finally, we generate the prediction of target segmentation masks sequence $\hat{\mathcal{S}}$ for the sounding objects in the frame sequence by convolving the fused multi-modal mask features $\mathcal{F}_m$ with the generated convolutional kernels ${G_{\text{kernel}}}$:

\begin{equation}
    \hat{\mathcal{S}} =  \mathcal{F}_\text{m} \star G_{\text{kernel}},
\end{equation}
where $\hat{\mathcal{S}}\in \mathbb{R}^{N_{q}\times T \times H_{lr}\times W_{lr} }$ are the predicted low-resolution segmentations for all queries.

\subsection{Training}

After acquiring segmentation masks for all the queries, 
we select the one that best refers to the sounding objects with bipartite matching strategy. Specifically, denoting the ground-truth segmentation mask as $y$ 
and the predicted segmentation mask from our model as $\hat{\mathcal{Y}}=\{\hat{y}_i\}_{i=1}^{N_q}$, we find the best matched segmentations with the lowest cost to the ground-truth:
\begin{equation}\label{eq:matching}
    i^* = \operatorname*{argmin}_{i\in \{1,\dots, N_q\}}\mathcal{C}(\hat{y}_{i},y),
\end{equation}
where $\mathcal{C}$ is the pair-wise cost using certain metrics;
and $i^*$ is the optimal query index. 
In addition to the segmentation head, we also append a binary classification head to map each query embedding to a single \textit{sounding score} that represents the possibility of the existence of sounding objects corresponding to the audio queries.

The cost metrics $\mathcal{C}$ in Eq.~\ref{eq:matching} consist of two parts: 
one for segmentation~($\mathcal{C}_{seg}$), that includes binary focal loss $\mathcal{C}_{focal}$ and DICE loss~\cite{milletari2016v} $\mathcal{C}_{dice}$,
and the other is binary classification for sounding objects presence~($\mathcal{C}_{sound}$). The total loss function between query $i$ and ground-truth can be written:
\begin{equation}
    \mathcal{C}(\hat{s}_i, s) = \lambda_{\text{dice}}\mathcal{C}_{\text{dice}} + \lambda_{\text{focal}} \mathcal{C}_{\text{focal}} + \lambda_{\text{sound}} \mathcal{C}_{\text{sound}},
\end{equation}
where  $\lambda_{\text{dice}}$, $\lambda_{\text{focal}}$ and $\lambda_{\text{sound}}$ are the weights to balance the costs.
The whole framework is trained by minimizing the total cost between $\hat{y}_{i^*}$ and ground-truth $y$.

\subsection{Inference}
At the inference stage, given a video frame sequence and its corresponding audios,
our model predicts the segmentation mask sequence for all query instances $\hat{\mathcal{Y}}=\{\hat{y}_i\}_{i=1}^{N_q}$, 
and corresponding sounding scores ${z}\in \{z_i \}_{i=1}^{N_q}$.
We pick the segmentation results with the highest sounding score $z$
and use bilinear sampling to match the resolution sizes of ground-truth segmentation masks.

\section{Experiment}\label{sec:exper}
\noindent \textbf{Datasets.}
We evaluate the proposed method on the AVSBench \cite{zhou2022avs} dataset with pixel-level audio-visual segmentation annotations.
This dataset consists of two subsets: the semi-supervised Single Sound Source Segmentation (S4) and the fully supervised Multiple Sound Source Segmentation (MS3).

\vspace{0.1cm}
\noindent \textbf{Metrics.}
We adopt the measures including $\mathcal{M}_{\mathcal{J}}$ (the mean Intersection-over-Union) and $\mathcal{M}_{\mathcal{F}}$ (F-score) following~\cite{zhou2022avs}.

\vspace{0.1cm}
\noindent \textbf{Implementation details.}
We freeze the parameters of visual and audio backbones.
We choose the Deformable Transformer~\cite{zhu2020deformable} as the transformer architecture in the framework. 
We choose AdamW~\cite{loshchilov2017decoupled} optimizer with initial learning being $10^{-4}$ and weight decay being  $5\times 10^{-4}$.
The models are trained for 50 epochs with single GeForce RTX 3090 and the training batch size is set to 8.

\begin{table}[hbt]
    \small
    \centering
    \caption{Comparison between different methods on two subsets of AVSBench dataset. }
    \begin{tabular}{ll|cccc}
    
    \toprule
    \multicolumn{2}{c}{Subsets}  &  \multicolumn{2}{c}{S4} & \multicolumn{2}{c}{MS3} \\
     \cmidrule(r){3-4} \cmidrule(r){5-6} \multicolumn{2}{c }{Methods}   &   $\mathcal{M}_{\mathcal{J}}$ &  $\mathcal{M}_{\mathcal{F}}$ & $\mathcal{M}_{\mathcal{J}}$ & $\mathcal{M}_{\mathcal{F}}$ \\
    \midrule
    \multirow{2}{*}{SSL} & LVS~\cite{chen2021localizing} &37.9 &.510 &29.5 &  .330   \\ 
     & MSSL~\cite{qian2020multiple} & 44.9  & .663 &26.1 &.363 \\  \cline{2-6}
     \multirow{2}{*}{VOS} & 3DC~\cite{mahadevan2020making} & 57.1 & .759  & 36.9 & .503    \\ 
     & SST~\cite{duke2021sstvos} & 66.3 & .801 & 42.6 & .572 \\   \cline{2-6}
     \multirow{2}{*}{SOD} & iGAN~\cite{mao2021transformer} &61.6 & .778 & 42.9& .544    \\ 
     & LGVT~\cite{zhang2021learning} & 74.9 & .873& 40.7 & .593 \\  \cline{2-6}
     \multirow{2}{*}{TPAVI~\cite{zhou2022avs}} & ResNet50 & 72.8 & .848 & 47.9 & .578    \\ 
     & PVT-v2 & 78.7 &.879 & 54.0 & .645 \\  \cline{2-6}
     \multirow{2}{*}{AuTR (Ours)} & ResNet50 & 75.0 & .852 & 49.4 & .612     \\ 
     & PVT-v2 & \textbf{80.4} & \textbf{.891} & \textbf{56.2} & \textbf{.672} \\
    %  & PVT-v2 + F.T. & - & -& \textbf{60.9} & \textbf{.725} \\ 
    \bottomrule
    \end{tabular}
    % \vspace{-3pt}
    % (``F.T." denotes finetuning based on the weights on S4)
    \label{tab:compare_with_sota}
\vspace{-0.7cm}
\end{table}

\subsection{Comparison with State-of-the-art Methods}

We compare AuTR against the current state-of-the-art method TPAVI~\cite{zhou2022avs}, which is based on the encoder-fusion-decoder framework. We also consider other related audio-visual methods, including sound source localization (SSL): LVS~\cite{chen2021localizing} and MSSL~\cite{qian2020multiple}, video object segmentation (VOS): 3DC~\cite{mahadevan2020making} and SST~\cite{duke2021sstvos}, and salient object detection (SOD): iGAN~\cite{mao2021transformer} and LGVT~\cite{zhang2021learning}.

As shown in Table~\ref{tab:compare_with_sota}, our proposed approach outperforms existing methods in both subsets. Even in the S4 subset, where TPAVI achieves high values on the $\mathcal{M}_{\mathcal{J}}$ metric (72.8 with ResNet50 and 78.7 with PVT-v2), our proposed method still shows improvements: 2.2 points higher with ResNet50 and 1.7 points higher with PVT-v2.
Additionally, we observe the performance on the MS3 subset is much worse than on S4 subset for all methods.
This may be due to the difficulty of multi-sound setting and the scarcity of training data.
The improvement of performance on MS3 subset will be elaborated in the next subsection.

\vspace{-5pt}
\subsection{Performance Improvement for Multiple Sound Sources}\label{sec:genera}
To further improve the performance for segmentation of multi-sound setting, we fine-tune the models pre-trained on the S4 subset for MS3. 
We conduct experiments using both TPAVI and our method separately.
As the results shown in Table~\ref{tab:s4fine}, by finetuning the weights of S4, the performance on the MS3 subset of our method is significantly boosted: on $\mathcal{M}_{\mathcal{J}}$ 6.69 \textit{improvement} with the ResNet50 backbone and 4.74 \textit{improvement} with PVT-v2.
This validates that our proposed audio-aware query-enhanced transformer can effectively extract and leverage the knowledge from single sound scenario and transfer it to multiple sound source scenes. 
On the contrary, for the TPAVI model, finetuning on model weights of the S4 subset causes noticeable performance degradation: on $\mathcal{M}_{\mathcal{J}}$ 3.56 \textit{drop} with ResNet50 backbone and 3.55 \textit{drop} with PVT-v2.
This performance degradation may be attributed to the fact that fusion-based models are prone to biasing towards single sound data, thus making it difficult to generalize to multi-sound scenario.
By comparing the results, our method shows strong generalization ability compared to TPAVI.

\begin{table}[]
\scriptsize
% \small
\setlength{\tabcolsep}{3.5pt}
\centering
\caption{Comparison between TPAVI and AuTR when fine-tuned on MS3 subset using the S4 subset trained models.  "F.T." denotes fine-tuning using the pre-trained weights on S4 subset.
}
\begin{tabular}{lccccc}
\toprule
\multirow{2}{*}{Method}  & \multirow{2}{*}{F.T.}  & \multicolumn{2}{c} {ResNet50} & \multicolumn{2}{c}{PVT-v2}  \\
\cmidrule(lr){3-4}\cmidrule(lr){5-6}
 &  & $\mathcal{M}_{\mathcal{J}}$ & $\mathcal{M}_{\mathcal{F}}$  & $\mathcal{M}_{\mathcal{J}}$ & $\mathcal{M}_{\mathcal{F}}$  \\
 \midrule
\multirow{2}{*}{TPAVI~\cite{zhou2022avs}} & $\times$ & 47.90  & .578 & 54.00 & .645 \\
& $\checkmark$ & 44.34 ($\downarrow$3.56)  & .583 ($\uparrow$0.005) & 51.45 ($\downarrow$ 3.55) & .671 ($\uparrow$.026 )\\
\midrule
\multirow{2}{*}{AuTR (Ours)} & $\times$  &49.41 & .612 & 56.21 & .672 \\
& $\checkmark$ & \textbf{56.00} ($\uparrow$6.59)  & \textbf{.660} ($\uparrow$0.048) & \textbf{60.95} ($\uparrow$ 4.74) &  \textbf{.725} ($\uparrow$ 0.053)\\
\bottomrule
\end{tabular}
% \vspace{-0.55cm}
\label{tab:s4fine}
\vspace{-0.5cm}
\end{table}

\vspace{-4pt}
\subsection{Open Set Audio Visual Segmentation}
To further evaluate the generalization of our method, we explore the task of open set segmentation. 
Specifically, we first randomly sample 11 categories of videos from the S4 subset for model training (seen), and then we test the trained models on the remaining 12 disjoint categories (unseen) without further training. We compare the open-set performance of AuTR with the TPAVI~\cite{zhou2022avs} method in Table~\ref{tab:openset}.
For seen categories, our model has better performance than TPAVI in both metrics.
When tested on the unseen categories, both methods have the performance drop.
E.g., TPAVI~\cite{zhou2022avs} drops from 75.62 to 55.86, our model drops from 77.56 to 66.22 on $\mathcal{M}_{\mathcal{J}}$.
In spite of this, AuTR still outperforms TPAVI~\cite{zhou2022avs} by 10.36 points with PVT-v2 on $\mathcal{M}_{\mathcal{J}}$.
In conclusion, our proposed AuTR architecture shows impressive generalisation performance compared with the encoder-fusion-decoder baseline TPAVI~\cite{zhou2022avs}.

\begin{table}[]
% \scriptsize
% \small
\centering
\setlength{\tabcolsep}{6pt}
\caption{Open set segmentation performance comparison between TPAVI~\cite{zhou2022avs} and our proposed AuTR.}
\begin{tabular}{llcccc}
\toprule
\multirow{2}{*}{Method}  & \multirow{2}{*}{Backbones}  & \multicolumn{2}{c} {Seen categories} & \multicolumn{2}{c}{Unseen categories}  \\
\cmidrule(lr){3-4}\cmidrule(lr){5-6}
 &  & $\mathcal{M}_{\mathcal{J}}$ & $\mathcal{M}_{\mathcal{F}}$  & $\mathcal{M}_{\mathcal{J}}$ & $\mathcal{M}_{\mathcal{F}}$  \\
 \midrule
\multirow{2}{*}{TPAVI~\cite{zhou2022avs}} & ResNet50 & 68.93  & .815 & 47.46 & .683 \\
& PVT-v2 & 75.62  & .862  & 55.86  & .719 \\
\midrule
\multirow{2}{*}{AuTR (Ours)} & ResNet50  &  70.46 & .817  & 51.15 & .675 \\
& PVT-v2 & \textbf{77.56} & \textbf{.865} & \textbf{66.22} & \textbf{.777} \\
\bottomrule
\end{tabular}
% \vspace{-0.55cm}
\label{tab:openset}
\vspace{-0.3cm}
\end{table}

% \vspace{-20pt}
\subsection{Ablation Analysis}
\begin{table}[]
% \small
\centering
\setlength{\tabcolsep}{13pt}
\caption{Ablation analysis of AuTR framework; w/o: without; AAQ: Audio-aware queries; DynConv: dynamic convolution.}
\begin{tabular}
{l c c c}
\toprule
Model &  w/o AAQ & w/o DynConv & AuTR  \\
\midrule
$\mathcal{M}_{\mathcal{J}}$ & 79.6  & 79.5  & 80.4 \\
$\mathcal{M}_{\mathcal{F}}$ & 0.882  &   0.890     & 0.891  \\
\bottomrule
\end{tabular}
\label{tab:ablation}
\vspace{-6pt}
\end{table}

We conduct ablation studies on the S4 subset to evaluate the effectiveness of the audio-aware queries and the dynamic convolution in the framework.
The results in Table~\ref{tab:ablation} show that without either of the two components the model performance drops, verifying the effectiveness of audio-aware queries and dynamic convolution module in the proposed framework.

\vspace{-5pt}
\subsection{Qualitative Examples}

\begin{figure}[!htb]
\centering
\vspace{-0.2cm}
\includegraphics[width=.98\linewidth]{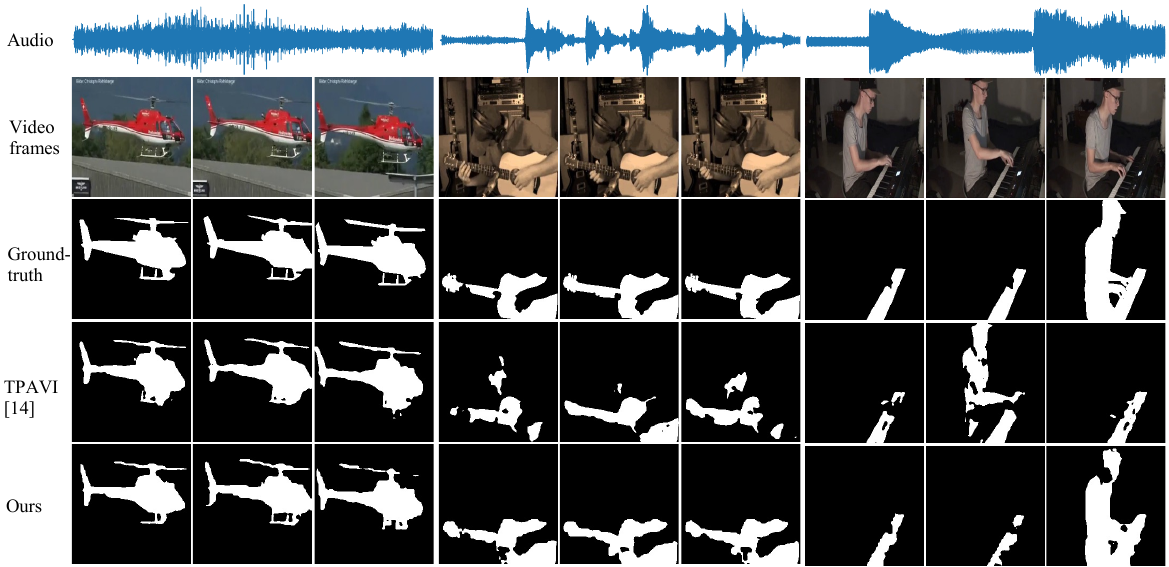}
\caption{Qualitative results of TPAVI~\cite{zhou2022avs} and AuTR on AVSBench test sets.}
\label{fig:qual-s4-ms3}
\vspace{-0.48cm}
\end{figure}

We present some segmentation results of TPAVI~\cite{zhou2022avs} and AuTR in~\cref{fig:qual-s4-ms3}.
It can be observed that our method can accurately localize and segment the sounding objects, which is closer to ground-truth. Additionally, in more challenging multi-sound scenarios, our method focuses on segmenting the target sounding objects while ignoring distracting objects. In contrast, TPAVI~\cite{zhou2022avs} tends to ignore audio cues and segments even silent yet salient objects in some frames.

\section{Conclusion}\label{sec:conclu}
We have introduced a highly effective multi-modal transformer framework for addressing the AVS problem.
Unlike traditional fusion-based methods, we leverage  transformer model to facilitate deep fusion and interaction of multi-modal features. Furthermore, we enhance the transformer decoder with audio-aware learnable queries that can explicitly help focus on sounding objects while suppressing salient yet silent objects.
Extensive experiments demonstrate that our approach achieves state-of-the-art performance and exhibits strong generalization ability.

% \clearpage
%%%%%%%%% REFERENCES
{\small
\bibliographystyle{IEEEbib}
\bibliography{refs}
}

\clearpage

\end{document}